%% file: BlR07.tex
\documentclass[conference,11pt,a4paper]{IEEEtran}
\usepackage{fancyvrb}
\usepackage{enumerate}
\usepackage{graphicx}
\usepackage{color} %
\usepackage{amsmath}
\usepackage{amscd}
\usepackage{balance}
\usepackage{msc}
\usepackage{cite}      %
\usepackage{graphicx}  %
\usepackage{url}       %
\usepackage{amsmath}   %
\begin{document}
\title{From Declarative Model to Solution: \\ 
Scheduling Scenario Synthesis}
\author{ \authorblockN{Bruno Bla\v{s}kovi\'{c}, Mirko Randi\'{c}}
   \authorblockA{University of Zagreb \\ Faculty~of~Electrical~Engineering~and~Computing } 
$\{bruno.blaskovic, mirko.randic\}@fer.hr$ }
\maketitle
\begin{abstract}
This paper presents deductive programming for scheduling scenario 
generation.  
Modeling for solution is achieved through program transformations.  
First, declarative   model for scheduling problem domain is  introduced.  
After that model is interpreted as scheduling domain   language and as predicate transition Petri net.  
Generated reachability tree presents search space with solutions.
At the end results are discussed and analyzed.
\end{abstract}
\section{Introduction}
Two general directions are under consideration in this paper. 
First, deductive programming will be used  
as methodology for solution to scheduling problem.
The second direction is  experience that improves protocol synthesis due to 
synergism between scheduling and deductive programming. 
Declarative programming is concerned about \textit{what} is to be done rather then \textit{how} is it implemented. 
Declarative model is interpreted and transformed to executable model.\\
In this paper word \textit{model} is frequently used. \\
\textit{Model} can present requirements, program, agent or behavior in pure mathematical way or through the program code. 
Nowadays, there are numerous formal methods, specification languages, model checking and theorem proving tools. 
Putting together different methods, tools and languages is obtained through model transformation, 
component composition software composition or similar methods.\\
This paper use different models, each of them is suitable for its particular purpose. 
Declaration part comes from language specialized for scheduling problem definition. 
Executable part is found in high level Petri net. 
Together, by means of model transformation solution to the problem is found. \\
This paper is structured as follows: 
before model translation between declarative and executable models two solutions are presented, 
one by means of \textit{Predicate Petri net} ($Pr/T$) in Section \ref{sec:prt} 
and the other by \textit{Planning Domain Definition Language} ($PDDL$) in Section \ref{sec:pddl}, respectively.\\
Working example is introduced in textual form in Section \ref{sec:example}. 
In Section \ref{sec:programming} unification between the $Pr/T$ and $PDDL$ model yielding 
translation between the models is introduced.
Section \ref{sec:scheduling} introduces metamodel as generalization of model transformations. \\
Experience from model translation and scheduling synthesis is used for scenario synthesis in Section \ref{sec:synthesis}. \\
Solution to scheduling problem as \textit{extended finite state machine} ($eFSM$) and 
$ITU-T$ \textit{message sequence diagram} ($MSC$) is in Section \ref{sec:solution}.\\
Final Sections of the paper bring related work (Section \ref{sec:related}) with some 
reflection regarding synthesis process  (in Section \ref{sec:scope}) as well as briefly recapitulate 
\textit{literate programming} methodology and \verb+noweb+ tool. \\
At the end in Section \ref{sec:conclusion} is conclusion with further research directions.\\
\section{Predicate Petri net}
\label{sec:prt}
In this Section \textit{Predicate Petri net} ($Pr/T$) solution is described. 
Textual problem from working example (Section \ref{sec:example}) is defined by ($Pr/T$) constructs and analyzed.
In following text \textit{working example} will be referenced as \verb+4ws1tob-problem+ shorter as \verb+4ws1tob+. \\
From the modeling point of view two models can be identified:
\begin{enumerate}[-]
\item mathematical model: $Pr/T$ is introduced as 6--tuple
\item program code that is input to $PrT$ tool for analysis
\end{enumerate}
Predicate-Transition Petri net definition is taken from \cite{prod:report-A26}. 
The tool  implementing $Pr/T$ \cite{prod:report-B11} has been derived following the same formal definition. 
$Pr/T$ is 6--tuple structure or mathematical $Pr/T$ model $(S,T,F,K,W,M_0)$ such that: 
\begin{enumerate}[\ ]
\item $S$ is the set of places,
\item $T$ is the set of transitions, $S \cap T = \emptyset$,
\item $F$ is the set of arcs, $F \subseteq (S \times T) \cup (T \times S)$,
\item $K$ is the capacity function, $K \in (S \to N_{\omega})$,
\item $W$ is the arc weight function, $W \in (F \to (N \setminus \{0\}))$, 
\item $M_0$ is initial marking (in initial state)  $M_0 \in \mathcal{M}$ where $\mathcal{M}$ is the set of markings (states), 
$\mathcal{M} = \{ M \in (S \to N) \mid \forall s \in S \quad M(s) \le K(s) \}$.
\end{enumerate}
$Pt/T$ tool used in this paper is PROD \cite{prod:report-B11}, \cite{prod:report-B13}. 
Analysis is performed by means of reachability tree generation. \\
Figure \ref{fig:prt}. represents programming model $Pr/T$ for \verb+4ws1tob-problem+ example:
\begin{enumerate}[-]
\item places (represented as circles) are sides of the "bridge",representing \textit{Safe} and \textit{Unsafe} part of the bridge.
\item transitions (represented as boxes) are actions or events (\textit{toSafe} and \textit{toUnsafe}) denoting "crossings": 
$e_S$ is event when $(s_i,s_j)$ are crossing from \textit{Unsafe} to \textit{Safe}, 
and $e_U$ is $(s_k)$ crossing from \textit{Safe} to \textit{Unsafe}, respectively
\item $<m_0,m_1,m_2,m_3>$ are markings 
\end{enumerate}
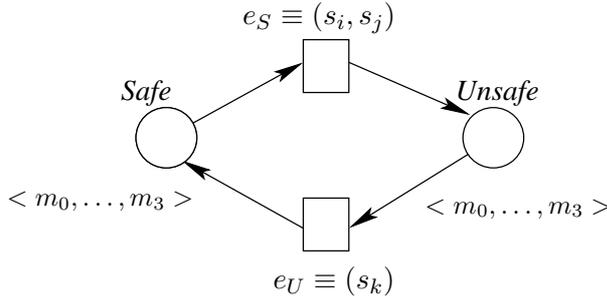
\begin{figure}[h]
\centering
\input{fig.prt.pstex_t}
    \caption{\small{Predicate Transition net for 4ws1tob problem (\textbf{Pr/T})}}
\label{fig:prt}
\end{figure}
There is no direct support for time in $Pr/T$ as well as in PROD program. 
They are fulfilled afterwords (subsection \ref{ssec:time}), by means of special program filter. 
More detailed description of $Pr/T$ in $PROD$ syntax shows that $Pr/T$ is also declarative \verb+4ws1tob+ 
problem description. 
In fact, graph structure from Fig.\ref{fig:prt} has program representation or program model that consists of:
\begin{enumerate}[(1)]
\item definitions: tokens in $Pr/T$ are of type integers, they are used to "carry" information about elapsed time,
\item places: \textit{Safe} and \textit{Unsafe}
\item transitions \textit{toUnsafe} and \textit{toSafe}
\end{enumerate}
Definitions are:
\begin{Verbatim}[fontsize=\small]
#define s1 10
#define s2 20
#define s3 25
#define torch 1
\end{Verbatim}
\textit{Unsafe} place has initial markings describing  "all soldiers are in \textit{Unsafe} place":
\begin{Verbatim}[fontsize=\small]
#place Unsafe \
  mk
   (<.s0.>+<.s1.>+
    <.s2.>+<.s3.>+<.torch.>) 
\end{Verbatim}
Each transition (\textit{toSafe} and \textit{toUnsafe}) implements previously mentioned events $e_S$ and $e_U$,
\verb+in+ is input place and \verb+out+ is output place, respectively:
\begin{Verbatim}[fontsize=\small]
#trans toSafe
in  {Unsafe: 
      <.x.>+<.y.>+<.torch.>;}
out {Safe:   
      <.x.>+<.y.>+<.torch.>;}

#trans toUnsafe
in  {Safe:   <.x.>+<.torch.>;}
out {Unsafe: <.x.>+<.torch.>;}
\end{Verbatim}
Goal is here expressed as computed tree logic (CTL) formula. 
Formula is used after reachability tree is generated. 
For that purpose separate program analyzer (\verb+probe+) is used. \\
Safe place will eventually have all tokens (or \textit{all soldier will be at safe side of the bridge}):
\begin{Verbatim}[fontsize=\small]
#define goal  
EventuallyOnSomeBranch
(safe == 
 <.1.>+<.5.>+<.10.>+<.20.>+<.25.>)
\end{Verbatim}
All paths (branches in CTL PROD terminology) with solutions are present. 
In order to decrease reachability tree timing constraints are separately calculated.
\subsection{Path \textit{filter}: time analysis}
\label{ssec:time}
Path filter selects only paths where goal--condition timing constraint holds:
$$
t_{elapsed} = \sum_{i=1}^{m}{t_i(e_i)} \le t_{max}
$$
where:\\
$t_i(e_i)$ -- event timing,\\
$m$ -- path length\\
One of such paths is presented in the Section \ref{sec:solution}.\\
As conclusion to this Section, experience from $Pr/T$ analysis can be applied to scheduling scenario generation: 
\begin{enumerate}
\item $Pr/T$ has mathematical or formal model expressed as 6--tuple with program representation--model in \verb+C-like+ syntax denoted as $\mathcal{M}_{PrT}(pd=4ws1tob)$
\item $Pr/T$ is also declarative model because it describes structure, analysis through reachability analysis establish $Pr/T$ as 
executable model.Executable model is denoted as $\mathcal{M}_{PROD}(pd=4ws1tob)$ ,
\item another declarative models that are established as a \textit{n--tuple} consisting of 
entities, predicates, events/actions and similar structure can be transformed to $Pr/T$.  
\end{enumerate}
\section{Planning Domain Description Language}
\label{sec:pddl}
\verb+PDDL+ (\textbf{P}lanning \textbf{D}omain \textbf{D}efinition \textbf{L}anguage) belongs to PDL  
(\textbf{P}roblem \textbf{D}omain \textbf{L}anguage) \cite{pddl:ipc5} class of languages. 
\verb+PDDL+ has syntax similar to \verb+Lisp+ and describes \textit{what} 
has to be done rather than \textit{how} is implemented.
That fact makes \verb+PDDL+ natural candidate for declarative Modeling. \\
\verb+PDDL+ main purpose is to serve as input language for many planning tools.  
In this paper \verb+PDDL+ is used as declarative input whose syntax is more general and intuitive 
and that can hide formal method from the user \cite{rushby00:disappear}.\\
Declarative \verb+PDDL+ model describing \verb+4ws1tob-problem+ is 5--tuple: 
\begin{equation*}
\begin{aligned}
  \mathcal{M}_{PDDL}&(pd=4ws1tob) =  \\
                    &(predicates, actions, \\
                    &objects, initial\_state, goal\_state)  \\
\end{aligned}%
\end{equation*}
where:
\begin{enumerate}[-]
\item objects: items of interest, for \verb+4ws1tob+ objects are objects=\{$s_0$, $s_1$, $s_2$, $s_3$\}
\item predicates: properties of objects, can be true or false, (example: Is $s_i$ in state $S$~?)
\item initial state(s): set of starting predicates formula (all $s_i$ in \textit{Unsafe})
\item goal state(s):  set of goal predicates formula (all $s_i$ in \textit{Safe})
\item actions (operators): "crossing" the bridge expressed through \textit{precondition} and \textit{effect} predicates
\end{enumerate}
In previous section (Sec.\ref{sec:prt}) place and state are 'words' with similar but in general case 
different meaning, because \verb+PDDL+ and $Pr/T$ languages have different semantic. 
In previous section (Sec.\ref{sec:prt}) $Pr/T$ has two models: mathematical and programming. 
\verb+PDDL+ has also two models, but both are expressed through \textit{Lisp}-like syntax.
\verb+PDDL+ can also serve as input to other planning and scheduling tools. \\
$\mathcal{M}_{PDDL}(pd=4ws1tob)$ problem is expressed through \verb+PDDL+ constructs. 
Each construct is \textit{Lisp}-like expression. 
Working example (\verb+4ws1tob+ from Sec. \ref{sec:example}) will be used to illustrate $\mathcal{M}_{PDDL}(pd=4ws1tob)$  
constructs. 
Now, we can say that \verb+PDDL+ program has the same syntax for mathematical and programming model. 
\verb+PDDL+ example starts with verbatim list of constructs:
\begin{Verbatim}[fontsize=\small]
(define (problem 4ws1tob01)
        (:domain 4ws1tob)
         (objects)
         (predicates)
         (initial_state)
         (goal_specification
         (actions_operators)))
\end{Verbatim}
Each construct will be described in more details.
\subsection{Objects}
Following notation from \cite{pvs:adtypes} types are introduced for each object: 
\begin{enumerate}[a)]
\item $s_0$, $s_1$,$s_2$, $s_3$ are objects of \textsl{type} sold,
\item \textit{torch} is object \textsl{type} torch,
\item \textit{Safe}, \textit{Unsafe} are objects of \textsl{type} place
\end{enumerate}
Objects in \verb+PDDL+ are not object from object oriented programming paradigm.
In \textit{Lisp}-like syntax objects are  
defined by term rewriting: \\
\begin{Verbatim}[fontsize=\small]
(objects) ::== 
   (:objects s0 s1 s2 s3 - sold 
     torch - torch
     Safe Unsafe - place)
\end{Verbatim}
Now object construct is \verb+PDDL+ executable, that means planning tools can execute it. 
Similarly other constructs are rewrote (or replaced) producing declarative specification.
\subsection{Predicates}
Predicates can be  used within other components. \\
\textit{Is token} \textit{sold} $s_i$ in \textit{place} $p_j$~? \textit{is expressed as}: 
\begin{Verbatim}[fontsize=\small]
   (:predicates
        (pl ?sold ?place))
\end{Verbatim}
\subsection{Initial states}
In initial state component all tokens ($s_0$, $s_1$, $s_2$, $s_3$) are in \textit{Safe} place and 
\textit{Unsafe} place is empty.
Timing parameters $t_i$ are set, too. 
Initial time is set as:
\begin{Verbatim}[fontsize=\small]
(:init 
 (= (t-elapsed) 0)
\end{Verbatim}
Initial state components are coded as follows: 
if token \verb+?x+ is in place \textit{Unsafe} than token \verb+?x+ is not in place \textit{Safe},
yielding following initial conditions:
\begin{Verbatim}[fontsize=\small]
 (pl s0  Unsafe) 
 (not (pl s0  Safe)) (= (ts s0) 5)
 (pl s1  Unsafe) 
 (not (pl s1  Safe)) (= (ts s1) 10)         
 (pl s2  Unsafe) 
 (not (pl s2  Safe)) (= (ts s2) 20)         
 (pl s3  Unsafe) 
 (not (pl s3  Safe)) (= (ts s3) 25)         
 (pl torch Unsafe))
\end{Verbatim}
Predicate (= (ts $s_i$) $t_i$) initialize crossing time for object $s_i$.\\
\subsection{Goal state}
Goal specification component is theorem about system behavior. 
If solution exists place \textit{Unsafe} is empty and 
all tokens of type \textit{sold} are in place \textit{Safe}.  
Solutions are found if goal is proved:
\begin{Verbatim}[fontsize=\small]
(:goal (and
  (pl s0 Safe) 
    (not (pl s0 Unsafe))
  (pl s1 Safe) 
    (not (pl s1 Unsafe))
  (pl s2 Safe) 
    (not (pl s2 Unsafe))
  (pl s3 Safe) 
    (not (pl s3 Unsafe))
  (pl torch Safe)
\end{Verbatim}
Goal has timing goal condition $t_{elapsed} \le 60$ expressed as:
\begin{Verbatim}[fontsize=\small]
  ((<= t-elapsed) 60)))
\end{Verbatim}
\subsection{Actions}
Action operators realize the following functionality:
\begin{enumerate}[a)]
\item two objects (or tokens) are transfered from \textit{Unsafe} to \textit{Safe},
time $t_{elapsed}$ incremented
\item single object (or token) is transfered from \textit{Safe} to \textit{Unsafe}, time $t_{elapsed}$ incremented
\item redundant token \textit{torch} is left in \verb+PDDL+ because implementation must support
silent--moves ($\epsilon$-actions)
\item parameters \verb+?x+ and \verb+?y+ are of type \textit{sold}
\end{enumerate}
Objects are used within \verb+PDDL+ terminology while tokens are used within $Pr/T$ terminology. 
Model transformations unifies objects and tokens, they will be mixed and used as synonyms.
Each action consist of \textit{preconditions} and \textit{effect}:
\begin{enumerate}[-]
\item \textit{Effect} is $e_S$ or $e_U$ event mentioned earlier in Fig.
\ref{fig:prt} Sec.\ref{sec:prt}.
\item \textit{Precondition} must hold in order an \textit{effect} takes place.
\item \textit{Preconditions} for \textit{toSafe} action are two tokens of type \textit{sold} in place
  \textit{Unsafe}.
\item \textit{Precondition} for \textit{toUnsafe} action  is token of type \textit{sold} in place \textit{Unsafe}.
\end{enumerate}
\subsubsection{\textit{toSafe} action}
~\\Event $e_S$ is realized with \textit{toSafe} action:
\begin{Verbatim}[fontsize=\small]
(:action toSafe :parameters (?x ?y)
 :precondition 
  (and (pl ?x Unsafe) (not (pl ?x Safe))
     (pl ?y Unsafe) (not (pl ?y Safe))
     (pl torch Unsafe) (not (pl torch Safe)))
\end{Verbatim}
\textit{Effect} should place chosen tokens in \textit{Safe} place:
\subsubsection{\textit{toUnsafe} action}
~\\Event $e_U$ is realized with \textit{toUnsafe} action:
\begin{Verbatim}[fontsize=\small]
:effect      
 (and 
  (pl ?x Safe)
    (not (pl ?x Unsafe))
  (pl ?y Safe)
    (not (pl ?y Unsafe))
\end{Verbatim}
and increment elapsed time $t_{elapsed}$:
\begin{Verbatim}[fontsize=\small]
 (+ (t-elapsed 
    (max (ts ?x)(ts ?y))))))
\end{Verbatim}
\textit{toUnsafe} action is similar to \textit{toSafe} action, 
single token of type \textit{sold} is going to \textit{Safe} place 
and token \verb+?x+ is removed from \textit{Safe} place and put into the 
\textit{Unsafe} place.
\textit{Precondition} with \textit{effect} is semantically equivalent to \verb+condition-event+ or 
\verb+Place-transition+ in Petri nets. 
That enables smooth model transition to non-colored Petri nets. 
\begin{Verbatim}[fontsize=\small]
(:action toUnsafe :parameters (?x)
:precondition
 (and
  (pl ?x Safe)    (not (pl ?x Unsafe))
  (pl torch Safe) (not (pl torch Unsafe)))

:effect
 (and
  (pl ?x Unsafe) (not (pl ?x Safe))
  (pl ?y Unsafe) (not (pl ?y Safe))
  (+ (t-elapsed (ts ?x)))))
\end{Verbatim}
\verb+PDDL+ described in this paper  produces the same results with \verb+(lpg) planning+ software. 
Program is executable after minor adjustments through software provided by \cite{pddl:ipc5} project. \\
Parameters ($N=4$, $K_S=2$, $K_U=1$, $t_{max}=60$) are preserved through transformation 
from $\mathcal{M}_{PROD}(pd=4ws1tob)$ 
to $\mathcal{M}_{PrT}(pd=4ws1tob)$.  \\
\section{Programming for solution}
\label{sec:programming}
In this paper intention is to derive executable model from deductive or declarative model. %
Terms \textit{deductive} and \textit{declarative} are used as synonyms although from the formal point of view 
it is not the same. \\
Intention is to define model ($\mathcal{M}(pd=4ws1tob)$) as executable without inventing yet another  
specialized Modeling or specification language. 
That opens  possibilities for reasoning about the model properties and consequently introduces validation 
in early development phase. \\
This hypothetical \verb+C+ program becomes deductive program. %
Deductive or declarative program must have implicitly defined algorithm that should deduce %
only from declarations and predicates output results. 
Such \verb+C+ program %
describes \textit{What} is done rather than \textit{How} is it done. \\
The same proposition holds for $\mathcal{M}(pd=4ws1tob)$ \verb+PDDL+ and $PrT$ models. 
We shall use shorter notation, $\mathcal{M}(pd)$ where $pd$ is always $pd=4ws1tob$.
$\mathcal{M}(pd)$ is focused on \textit{What} is to be done rather then \textit{How} is it done. 
Natural candidates for the model $\mathcal{M}(pd)$ translation are Prototype Verification System (\textbf{PVS}) 
\cite{pvs:adtypes}, term--rewriting systems and  \textit{Lisp} family of languages. 
Our solutions uses \textit{Lisp} like languages.\\ 
\textit{Modeling for solution} effect
is achieved through the following model transformations presented as commutative diagram in Fig.\ref{fig:cd}. 
Such approach verifies proof--of--concept  through model transformation experiments.\\
$TR_j$ and $TR_k$ are program transformation routine. 
In practical solution $TR_j$ and $TR_k$ will be realized through the metamodel concept: 
deductive will be interpreted through metamodel, metamodel is translated to executive model afterwords. 
In this paper direct model translation is used. Metamodel facilities are introduced in Section \ref{sec:scheduling}.
\begin{figure}[h]
  $$
\begin{CD}
  \mathcal{M}(pd) @>TR_j>> \mathcal{PDDL}_i @>TR_k>> \mathcal{PN}_p \\
\end{CD}
$$
\centering
\caption{\small{Commutative diagram for model transformations}}
\label{fig:cd}
\end{figure}
Each transformation between models $\mathcal{M}(pd)$ require parser, because model transformation is program transformation.
In order to avoid parser development following facts are considered:
\begin{enumerate}
\item mathematical models for \verb+PDDL+ is 5--tuple, introduced with lisp syntax,
\item mathematical models for $PrT$ is 6--tuple, expressed as mathematical text, not as programming language
\item $PROD$ program is in\verb+C+--like syntax and presents instantiation of $PrT$ 
\end{enumerate}
\verb+PROD+ program describing $PrT$ is coded in Lisp like constructs:
\begin{Verbatim}[fontsize=\small]
 #trans toSafe 
  in {Unsafe: <.x.>+<.y.>+<.torch.>;} 
  out {Safe:  <.x.>+<.y.>+<.torch.>;}
\end{Verbatim}
becomes Lisp $PROD$ or $lPROD$:
\begin{Verbatim}[fontsize=\small]
  (:trans toSafe 
    :parameters (?x ?y ?torch)
       :in  (Unsafe ?x ?y ?torch) 
       :out (Safe:   ?x ?y ?torch)
\end{Verbatim}
Note the similarity between \verb+PDDL+ \verb+:action+ construct and $lPROD$ \verb+:trans+ construct.\\

\subsection{Unification}
There is set of mappings between $PDDL$ and $lPROD$:
\begin{enumerate}[-]
\item $:init \longleftrightarrow$ initial-marking
\item $:goal \longleftrightarrow$ final-marking
\item $:action \longleftrightarrow$ \verb+#trans+
\item $:objects \longleftrightarrow$ \verb+<.tokens.>+     
\end{enumerate}
Translation between $PDDL$ and $lPROD$ is straightforward: the set of mappings unify \verb+PDDL+ and $lPROD$.
\section{Example scenario}
\label{sec:example}
This example belongs to the set of "toy--problems" used in experiments during algorithm testing.
\subsection{Textual scheduling problem definition}
\label{ssec:textual}
Working example is simple scheduling problem taken from \cite{noweb:4ws1tob}, 
listed verbatim:
\begin{quote}
  Four soldiers who are heavily injured, try to flee to their home
  land.  The enemy is chasing them and in the middle of the night they
  arrive at a bridge that spans a river which is the border between
  the two countries at war.  The bridge has been damaged and can only
  carry two soldiers at a time.  Furthermore, several land mines have
  been placed on the bridge and a torch is needed to sidestep all the
  mines.  The enemy is on their tail, so the soldiers know that they
  have only 60 minutes to cross the bridge.  The soldiers only have a
  single torch and they are not equally injured.  The following table
  lists the crossing times (one-way!)
  for each of the soldiers:
  \begin{enumerate}[-]
  \item soldier $S_0$ 5 minutes 
  \item soldier $S_1$ 10 minutes 
  \item soldier $S_2$ 20 minutes 
  \item soldier $S_3$ 25 minutes 
  \end{enumerate}
Does a schedule exist which gets all four soldiers 
to the safe side within 60 minutes?
\end{quote}
\section{Scheduling domain metamodel}
\label{sec:scheduling}
abstract model here -- no metamodel needed here
\label{sec:fm}
Scheduling domain metamodel is derived from \verb+PDDL+ model. 
Metamodel supports constructs from \textit{type} theory, concurrency theory as well as process algebras. 

After the analysis of the text from Section \ref{sec:example} the following list of \verb+constructs+ are introduced:
parameters, entities, predicates, events, traces, initial--conditions, goal--conditions, operators and constraints.\\
In the next step each construct is described through \textit{Lisp}-like constructs:
\begin{Verbatim}[fontsize=\small]
(def-abstract-semantic-net 
    "metametamodel" 
 (problem-domain 4ws1tob)
  (parameters construct)
  (entities construct) 
  (predicates construct)
  (events construct) 
  (traces construct)  
  (initial-conditions construct)
  (goal-conditions construct) 
  (operators construct) 
  (constraints construct))
\end{Verbatim}
Parameters are data of types \textit{integer} or \textit{real}: 
\begin{enumerate}[(1)]
\item number of soldiers $n=4$
\item \ldots \verb+carry two soldiers+ (to \textit{safe} side) $K_S=2$ , 
and $K_U=1$ (to \textit{unsafe}) side.
\item \ldots\verb+cross times+: $t_0=5$, $t_1=10$, $t_2=20$, $t_3=25$,
\item \ldots have only \verb+60 min.+ to cross the bridge 
\end{enumerate}
Torch is not considered here because it has not influence on model behavior. 
Next models ($\mathcal{PDDL}$, $hl\mathcal{PN}$) can include it but that is not necessary. 
\begin{Verbatim}[fontsize=\small]
(def-parameters
 (n 4 int) 
 (KS 2 int) 
 (KU 1 int)
 (t0 5 real)
 (t1 10 real)
 (t3 20 real)
 (t4 25 real)
 (t-max 60 real))
\end{Verbatim}
Entities are two sets: 
one set are variables and the other are values. 
Set $A$ is describing dynamic behavior of the model: in each execution step 
values from set $A_S$ are assigned to variables from $S$.
\begin{enumerate}
\item set of soldiers: $s_0 \ldots s_3$ of \textit{type} \verb+sold+
\item set describing sides of the bridge. 
They are introduced as \textit{places} (\textit{Safe} side and \textit{Unsafe} side) of \textit{type} \verb+place+.
\end{enumerate}
Unknown parameter is denoted as $?A$ or $?AS$.
\begin{Verbatim}[fontsize=\small]
(entity (A (s0 s1 s2 s3 s4))
(entity (AS (Safe Unsafe))
\end{Verbatim}
Predicates answers the question:
\begin{enumerate}[(1)]
\item Where is $s_i$~?
\item Is $s_i$ in side \textit{Safe} or \textit{Unsafe}
\end{enumerate}
\begin{Verbatim}[fontsize=\small]
(pred atPlace ?A)
(pred ?A ?AS)   
\end{Verbatim} 
There are two atomic events: 
\begin{enumerate}[(1)]
\item two soldiers $s_i$ and $s_j$ are going to \textit{Safe} side
\item single soldier $s_i$ is going to \textit{Unsafe} side
\item event has duration time $t_i$
\end{enumerate}
\begin{Verbatim}[fontsize=\small]
(eS (Unsafe (?x ?y) Safe)
    (time (max (?tx ?ty))))

(eU (Safe ?x Unsafe)
    (time (?tx )))
\end{Verbatim}
If there is solution for this problem traces should be of finite length $r$ coded as finite length vector, 
such that \verb+?e+ is \verb+?eS+ or \verb+?eU+ event of duration \verb+?total-time+.
This model has no built--in infinite traces.
\begin{Verbatim}[fontsize=\small]
(Er (foreach r ?e) ?total-time)
\end{Verbatim}
Initially all $s_i$ are on \textit{Unsafe} place. 
This model has no time counter 
  (initial-condition 
   (A (atUnsafe atUnsafe 
       atUnsafe atUnsafe)))
At the end all soldiers must be within $60$ minutes in safe side:
\begin{Verbatim}[fontsize=\small]
(goal-condition 
   (A (atSafe atSafe atSafe atSafe)) 
   (<= total-time t-max))
\end{Verbatim}
\textit{Operators} and \textit{constraints} constructs serve as additional model input 
in complex situations where $\mathcal{M}(pd)$ is profiled recursively.\\[-.2cm]
\section{Solution}
\label{sec:solution}
There are 16 paths from total of 824 paths where timing condition $t_{elapsed} \le 60\, min$ holds. 
As an example one path is presented:  \\
\begin{center}
\begin{Verbatim}[fontsize=\small]
PATH 33
Node 0: transition toSafe 
  x = 5  y = 10
Node 1: transition toUnsafe
  x = 5
Node 7: transition toSafe
  x = 25  y = 20
Node 13: transition toUnsafe
  x = 10
Node 20: transition toSafe
  x = 10  y = 5
Node 21
\end{Verbatim}
\end{center}
\verb+Node 0,1,5,13+ \ldots  are nodes from reachability tree. 
Variable $x$ and $y$ are crossing times, for \textit{toSafe} transition or $e_S$ event crossing time is 
$max(x,y)$. 

\subsection{Visualization: $e\mathcal{FSM}$}
\label{ssec:visualisation}
\begin{figure}[tbh]
  \centering
\input{fig.solution.pstex_t}
    \caption{\small{solution as real--time program (\textit{eFSM})}}
\label{fig:eFSM}
\end{figure}
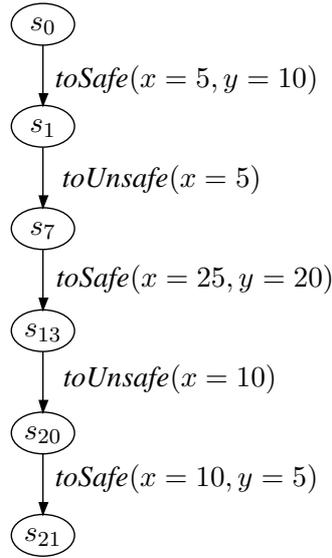

Fig.\ref{fig:eFSM} visualize \cite{graphviz:main} solution in the form of extended Finite State Machine (\verb+eFSM+). 
Next step can transform \verb+eFSM+ into the input language for analysis tool. 
Another possibility is to generate skeleton code in \verb+C+ or \verb+java+ programs. 
\subsection{MSC solution}
\label{ssec:msc}
\begin{figure}[h]
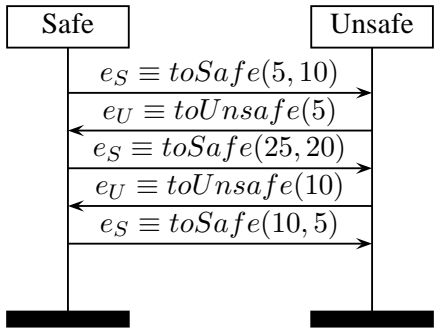

   \centering
\begin{msc}{4ws1tob.solution}
\drawframe{no}

  \declinst{a}{}{Safe}
  \declinst{b}{}{Unsafe}       

   \mess{$e_S \equiv toSafe (5,10) $}{a}{b}
\nextlevel

   \mess{$e_U \equiv toUnsafe (5) $}{b}{a}  
\nextlevel
   \mess{$e_S \equiv toSafe (25,20) $}{a}{b}
\nextlevel

   \mess{$e_U \equiv toUnsafe (10) $}{b}{a}  
\nextlevel
   \mess{$e_S \equiv toSafe (10,5) $}{a}{b}
\nextlevel

\end{msc}
     \caption{\small{solution as Message Sequence Diagram(\textbf{MSC})}}
 \label{fig:MSC}
 \end{figure}

Message sequence charts \cite{itu:Z120} is another form that can visualize solution (Fig.\ref{fig:MSC}). 
Even the more MSC can be used as source for another set of translations into the statecharts, SDL diagrams \ldots \\

\setlength{\instdist}{4cm}
\section{Synthesis and scheduling}
\label{sec:synthesis}
How can \textit{planning} and \textit{scheduling} methods influence synthesis?\\
Synergy effect between scheduling and synthesis opens possibilities for interpreting (\verb+eFSM+) in various ways. 
Each of them benefits through skeleton or templates for \verb+code+, \verb+scripts+ or \verb+architecture+ definition. 
During requirement phase many different scenarios can be automatically generated. 
We found that "side--effects" or parts of 
scenarios--specifications that are less obvious but present are reduced. \\
Decidability and computability of our approach is not optimal for big examples because of state explosion problem.
Heuristic scheduling algorithms are of little practical importance for synthesis problem. 
Different problem will in most cases have different declarative program. 
For that purpose Petri net reachability algorithms will be replaced with satisfiability (SAT) algorithms. \\
Some interpretation of solutions are:
\begin{enumerate}[a)]
\item protocol synthesis
\item \verb+SDL+ process: skeleton of program SDL can be generated and used by designer 
\item \verb+MSC+ skeleton processes can be composed in system. Overall behavior is analyzed as early as in requirement phase
\item parallel program job scheduling: an experiment for dynamic job allocation 
for parallel program is planned using proposed methodology
\end{enumerate}
Besides mentioned interpretations other possibilities are:
\begin{enumerate}[a)]
\item real--time system job scheduling 
\item control software synchronization 
\item performance prediction 
\item ontology definition concept analysis 
\item system maintenance
\item object and methods definition and optimization 
\end{enumerate}
Proposed approach introduces methodology for identifying and minimization of states in \verb+FSM+ like models of
concurrent reactive systems. \\
\section{Scope and motivation}
\label{sec:scope}
There are various approaches for synthesis problem, the most significant are: 
\begin{enumerate}[(a)]
 \item the temporal logic formula describes the system. Synchronization part of the system is derived from temporal formula. 
This can be, roughly speaking, interpreted as reverse model checking.
 \item from formal service specification to protocol specification. Formal description is transformed into protocol specification. 
Even the more, in most cases specification is executable enabling verification, simulation and analysis \ldots
 \item extended finite state machine is constructed from executable traces. 
Traces are sequences of messages, signals or sequence of events. 
They can be defined by designer, in this paper intention is to provide traces automatically to the designer.
\end{enumerate}
Our approach introduce traces or more preciously event traces as declaration for desired system behavior. 
An event describe \textit{crossing the bridge} 
(example from Section \ref{sec:example}. %
Sequence of events define trace. 
The set of all traces represents search space where solution should be found taking timing 
constraints into the considerations. \\
In this paper traces are interpreted as Petri net reachability tree paths. 
Reachability tree paths and traces describe the same behavior model.\\
If various models $\mathcal{M}$ have same behavior model then they can be transformed. 
Transformation is mapping of constructs between models, before mapping constructs are unified.\\
Only paths with desired property (crossing time limit) are solution paths. \\
Another question is how to only generate traces that are solution i.e. to avoid state--space combinatorial explosion.  
Declarative meta--model  has no knowledge about it. \\
Modeling for solution has three steps:
\begin{enumerate}[(i)]
\item model definition %
\item translation to domain specific language (in our case \verb+PDDL - scheduling&planning+ language). 
\verb+PDDL+ can be used as input for scheduling planning tools.
\item translation to high level Petri net for analysis and solution finding.
\end{enumerate}
It is obvious fact that \textit{model--for--solution} can start and find solution from 
step \textit{(ii)} or step \textit{(iii)} without the model. %
In complex situations, when system is not formally described, where constraints and assertions about the system 
are contradictory, unknown,  unclear or unspecified such model--mixing %
proves its value. 
Another motivation is to give  designer or modeler support to--play--with with  %
different tools and approaches in order to achieve desired quality of solution.
Formal approaches explore the benefits and experience from automatic deductive programming, program transformation as well as literate programming.  \\
Previous work were focused on synthesis as component composition: smaller architectural parts or 
system  blocks were composed into the target system. 
Such approach has usable results for component based  architectures like services definition 
within the intelligent networks as found in numerous $ITU-T$ recommendations. 
Later on, working example problem is solved with high level Petri net in a way close to approach described in Sec.\ref{sec:prt}. 
As a consequence, scheduling scenario interpreted as MSC scenario yields another synthesis approach. 
Such interpretation can produce MSC chart as solution or synthesize executable specification by means of scheduling methodology. 
Modeling for solution follows experiments towards synthesis of scenarios and its translation to finite state 
machine based systems like statecharts or ITU-T SDL language.
This paper also try to address such question through reachability tree analysis of scheduling solver. 
Results and methodology from another research field (planning and scheduling) are exercised,  yielding 
synergistic effect on protocol or concurrent reactive system design. 
Experience shows that scheduling problem generalization and  synthesis issues can benefit from each other.\\
In \cite{bla:contel03} Modeling framework suitable for experiments is introduced. 
Modeling framework consists of several levels, each level describes position in model hierarchy, 
from the most abstract level on the top to implementation level at the bottom. 
Within each level components are introduced (traditionally called $\mathcal{ECP}$ 
(\textbf{E}lementary \textbf{C}ommunicating \textbf{P}rocesses) as black boxes 
that enables program, tools or even models inter working. \\
%
\section{Related work}
\label{sec:related}
In \cite{psy:msc-room} synthesis is described as message sequence chart (MSC) translation into the Real-time Object Oriented Model 
(ROOM). 
After that, designer can use ROOM model for simulation as well as other purposes. 
Formals description technique (MSC) describing system architecture and behavior is interpreted as executable model. 
MSC serves as \textit{top--level--model} which can be analyzed, simulated and implemented. \\
Functional specification of the problem and temporal logic yields state--based automaton as 
solution for elevator problem \cite{psy:ursu}. 
Satisfiability analysis generates synchronization part of the system. \\
Synthesis of behavior models from scenarios is introduced in \cite{icse01:uchitel} and \cite{uchitel04:arch}.\\
\verb+PROMELA+ model serves as input of \verb+spin+ protocol verifier from \cite{noweb:4ws1tob}. 
Results are presented through MSC diagrams. \\
Results from mathematical description with process algebra and concurrency 
presented in \cite{bla:scibook2006} are used for further development of metamodel described in Section \ref{sec:scheduling}. 
Another metamodel comes from \cite{eclipse:km3}.
Model transformation routines are developed  by means of \verb+noweb+ literate programming tool.
Literate programing discipline has been introduced by D.E.~Knuth  
\ldots \textit{instead of imagining that our main task is to instruct a computer what to do, let us concentrate
rather on explaining to human beings what we want a computer to do.} \\
There are many literate programming supporting tools \cite{noweb:smith-towards} providing human readable files 
that incorporate documentation and source code into the single file. 
In this paper all sections illustrating concepts and constructs (functional style programs) 
are produced with literate programming tool \verb+noweb+ \cite{noweb:ram-simplified}, \cite{noweb:linux-journal}. 
\section{Conclusion and further work}
\label{sec:conclusion}
Synergy effect between planning, scheduling and synthesis can improve design process. 
There are no universal approach for synthesis problem so only narrow  problem domains are possible to solve 
with difficulties regarding NP-hard algorithms and undecidable problems. 
This papers describe proof--of--concept rather then industrial strength approach. \\
Pros (+) and cons (-) can be summarized as follows:
\begin{enumerate}[]
\item (+)~synergism between synthesis and scheduling planning: all ready developed routines for scheduling have been 
adopted and used
\item (-)~state explosion: Petri net can produce unmanageable  reachability tree size. 
Reachability analysis tool support is designed for model checking. Some scheduling issues are unsuitable 
for model--checking technology
\item (-)~narrow problem domain: declarative model requires significant changes with small domain change
\item (-)~small scale problems: synthesized components are sometimes easier to handle by hand
\item (+)~proof of concept: model transformation is usable programming paradigm
\item (-)~complex theoretical background: designer should have deep understanding of all models and translation process
\item (+)~solution for critical applications: mission critical software can be developed in this way yielding stable solutions
\item (+)~open research platform: modifications and updating to new algorithms
\item (+)~interworking of different paradigms and formal methods 
\end{enumerate}
Further work will (1) use satisfiability algorithms 
and (2) explore formal methods interworking. 
Synthesis method should serve as testbed for formal languages semantic analysis and executable specification languages. 
\end{document}

%% file: fig.prt.pstex_t
\begin{picture}(0,0)%
\includegraphics{fig.prt.pstex}%
\end{picture}%
\setlength{\unitlength}{4144sp}%
\begingroup\makeatletter\ifx\SetFigFontNFSS\undefined%
\gdef\SetFigFontNFSS#1#2{%
  \fontsize{#1}{#2pt}%
  \selectfont}%
\fi\endgroup%
\begin{picture}(3083,1819)(-14,-989)
\put(2656,209){\makebox(0,0)[lb]{\smash{{\SetFigFontNFSS{11}{13.2}{\color[rgb]{0,0,0}\textit{Unsafe}}%
}}}}
\put(676,209){\makebox(0,0)[lb]{\smash{{\SetFigFontNFSS{11}{13.2}{\color[rgb]{0,0,0}\textit{Safe}}%
}}}}
\put(1576,-916){\makebox(0,0)[lb]{\smash{{\SetFigFontNFSS{11}{13.2}{\color[rgb]{0,0,0}$e_U \equiv(s_k)$}%
}}}}
\put(1396,659){\makebox(0,0)[lb]{\smash{{\SetFigFontNFSS{11}{13.2}{\color[rgb]{0,0,0}$e_S \equiv (s_i,s_j)$}%
}}}}
\put(  1,-421){\makebox(0,0)[lb]{\smash{{\SetFigFontNFSS{11}{13.2}{\color[rgb]{0,0,0}\small$<m_0,\ldots,m_3>$}%
}}}}
\put(2476,-466){\makebox(0,0)[lb]{\smash{{\SetFigFontNFSS{11}{13.2}{\color[rgb]{0,0,0}\small$<m_0,\ldots,m_3>$}%
}}}}
\end{picture}%

%% file: fig.solution.pstex_t
\begin{picture}(0,0)%
\includegraphics{fig.solution.pstex}%
\end{picture}%
%
%
\setlength{\unitlength}{4558sp}%
\begingroup\makeatletter\ifx\SetFigFontNFSS\undefined%
\gdef\SetFigFontNFSS#1#2{%
  \fontsize{#1}{#2pt}%
  \selectfont}%
\fi\endgroup%
\begin{picture}(1013,3012)(-7,-1890)
\put(171,417){\makebox(0,0)[b]{\smash{{\SetFigFontNFSS{11}{13.2}{\color[rgb]{0,0,0}$s_1$}%
}}}}
\put(946,659){\makebox(0,0)[b]{\smash{{\SetFigFontNFSS{11}{13.2}{\color[rgb]{0,0,0}\textit{toSafe}$(x=5,y=10)$}%
}}}}
\put(171,971){\makebox(0,0)[b]{\smash{{\SetFigFontNFSS{11}{13.2}{\color[rgb]{0,0,0}$s_0$}%
}}}}
\put(991,-421){\makebox(0,0)[b]{\smash{{\SetFigFontNFSS{11}{13.2}{\color[rgb]{0,0,0}\textit{toSafe}$(x=25,y=20)$}%
}}}}
\put(856,-961){\makebox(0,0)[b]{\smash{{\SetFigFontNFSS{11}{13.2}{\color[rgb]{0,0,0}\textit{toUnsafe}$(x=10)$}%
}}}}
\put(946,-1501){\makebox(0,0)[b]{\smash{{\SetFigFontNFSS{11}{13.2}{\color[rgb]{0,0,0}\textit{toSafe}$(x=10,y=5)$}%
}}}}
\put(171,-1801){\makebox(0,0)[b]{\smash{{\SetFigFontNFSS{11}{13.2}{\color[rgb]{0,0,0}$s_{21}$}%
}}}}
\put(171,-1247){\makebox(0,0)[b]{\smash{{\SetFigFontNFSS{11}{13.2}{\color[rgb]{0,0,0}$s_{20}$}%
}}}}
\put(171,-692){\makebox(0,0)[b]{\smash{{\SetFigFontNFSS{11}{13.2}{\color[rgb]{0,0,0}$s_{13}$}%
}}}}
\put(171,-137){\makebox(0,0)[b]{\smash{{\SetFigFontNFSS{11}{13.2}{\color[rgb]{0,0,0}$s_7$}%
}}}}
\put(811,119){\makebox(0,0)[b]{\smash{{\SetFigFontNFSS{11}{13.2}{\color[rgb]{0,0,0}\textit{toUnsafe}$(x=5)$}%
}}}}
\end{picture}%